\documentclass[sigconf, screen]{acmart}

\copyrightyear{2024}
\acmYear{2024}
\setcopyright{rightsretained}
\acmConference[CCS '24]{Proceedings of the 2024 ACM SIGSAC Conference on Computer and Communications Security}{October 14--18, 2024}{Salt Lake City, UT, USA}
\acmBooktitle{Proceedings of the 2024 ACM SIGSAC Conference on Computer and Communications Security (CCS '24), October 14--18, 2024, Salt Lake City, UT, USA}
\acmDOI{10.1145/3658644.3690229}
\acmISBN{979-8-4007-0636-3/24/10}

\makeatletter
\gdef\@copyrightpermission{
  \begin{minipage}{0.3\columnwidth}
   \href{https://creativecommons.org/licenses/by/4.0/}{\includegraphics[width=0.90\textwidth]{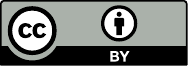}}
  \end{minipage}\hfill
  \begin{minipage}{0.7\columnwidth}
   \href{https://creativecommons.org/licenses/by/4.0/}{This work is licensed under a Creative Commons Attribution International 4.0 License.}
  \end{minipage}
  \vspace{5pt}
}
\makeatother

\usepackage{graphicx}
\usepackage{enumitem}%no
\usepackage{fancyvrb}
\usepackage{listings}
\usepackage{amsthm}
\usepackage{multirow}
\usepackage{bm}
\usepackage{textcomp}
\usepackage{tcolorbox}%no
\usepackage{colortbl}%no
\usepackage[ruled,vlined,linesnumbered]{algorithm2e}
\usepackage{algpseudocode} %no
\usepackage{bbding}
\newcommand{\Toolname}{\textsc{$\mathcal{E}$fuzz}}
\newcommand{\tool}{{\sc \Toolname}\xspace}
\newcommand{\toolBold}{{\sc\bfseries \Toolname}\xspace}
\newcommand{\klee}{{\sc KLEE}\xspace}
\newcommand{\aflnet}{{\sc AFLNet}\xspace}
\newcommand{\aflnetBold}{{\sc\bfseries AFLNet}\xspace}
\newcommand{\afl}{{\sc AFL}\xspace}
\newcommand{\aflplus}{{\sc AFL++}\xspace}

\newcommand{\epatch}{{\sc E9Patch}\xspace}

\newcommand{\eafl}{{\sc E9AFL}\xspace}

\newcommand{\nyxnet}{{\sc Nyx-Net}\xspace}
\newcommand{\nyxnetBold}{{\sc\bfseries Nyx-Net}\xspace}

\newcommand{\profuzzbench}{{\sc ProFuzzBench}\xspace}
\newcommand{\profuzzbenchBold}{{\sc\bfseries ProFuzzBench}\xspace}
\newcommand{\sgfuzz}{{\sc SGFuzz}\xspace}
\newcommand{\nsfuzz}{{\sc NSFuzz}\xspace}
\newcommand{\ijon}{{\sc IJON}\xspace}
\newcommand{\stateafl}{{\sc StateAFL}\xspace}
\newcommand{\snapfuzz}{{\sc SnapFuzz}\xspace}

\newcommand{\efone}{{\sc EF1}\xspace}
\newcommand{\efoneBold}{{\sc\bfseries EF1}\xspace}
\newcommand{\eftwo}{{\sc EF2}\xspace}
\newcommand{\eftwoBold}{{\sc\bfseries EF2}\xspace}

\definecolor{deepblue}{rgb}{0,0,0.5}
\definecolor{deepred}{rgb}{0.6,0,0}
\definecolor{deepgreen}{rgb}{0,0.5,0}
\definecolor{halfgray}{gray}{0.55}
\definecolor{ipythonframe}{RGB}{207, 207, 207}
\definecolor[named]{ACMBlue}{cmyk}{1,0.1,0,0.1}
\definecolor[named]{ACMYellow}{cmyk}{0,0.16,1,0}
\definecolor[named]{ACMOrange}{cmyk}{0,0.42,1,0.01}
\definecolor[named]{ACMRed}{cmyk}{0,0.90,0.86,0}
\definecolor[named]{ACMLightBlue}{cmyk}{0.49,0.01,0,0}
\definecolor[named]{ACMGreen}{cmyk}{0.20,0,1,0.19}
\definecolor[named]{ACMPurple}{cmyk}{0.55,1,0,0.15}
\definecolor[named]{ACMDarkBlue}{cmyk}{1,0.58,0,0.21}

\newcommand{\ignore}[1]{}

\newcommand*{\RN}[1]{\uppercase\expandafter{\romannumeral #1\relax}}
\newcommand*{\rn}[1]{\lowercase\expandafter{\romannumeral #1\relax}}

\definecolor{shadecolor}{gray}{1.00}
\definecolor{ddarkgray}{gray}{0.75}
\definecolor{darkgray}{gray}{0.30}
\definecolor{light-gray}{gray}{0.87}

\hypersetup{colorlinks,
  linkcolor=ACMDarkBlue,
  citecolor=ACMPurple,
  urlcolor=ACMDarkBlue,
  filecolor=ACMDarkBlue}

\newcommand{\etc}{\emph{etc.}\xspace}

\newcommand{\ie}{i.e.,\xspace}

\newcommand{\eg}{e.g.,\xspace}

\newcommand{\ccc}[1]{\raisebox{.9pt}{\textcircled{\raisebox{-.9pt}{\small #1}}}}

\theoremstyle{definition}

\definecolor{mycolor}{rgb}{0.122, 0.435, 0.698}
\newcommand{\result}[1]{%
\begin{tcolorbox}[colframe=mycolor,boxrule=0.5pt,arc=4pt,
      left=6pt,right=6pt,top=6pt,bottom=6pt,boxsep=0pt,width=\columnwidth]%
      {#1}
\end{tcolorbox}%
}

\definecolor{keyword}{rgb}{0.0, 0.50, 0}
\definecolor{comment}{rgb}{0, 0, 0.50}
\definecolor{const}{rgb}{0.6,0,0}
\definecolor{syscall}{rgb}{0.4,0.4,0}
\definecolor{background}{rgb}{0.95,0.95,0.95}

\newcommand{\crashingSeeds}{$\mathcal{C}_{\scalebox{.5}{\textit{\XSolidBrush}}}$\xspace}

\newcommand{\MyComment}[1]{\Comment{\colorbox{gray!30}{#1}}}

\begin{document}

\title{Program Environment Fuzzing}

\author{Ruijie Meng}
\affiliation{%
  \institution{National University of Singapore}
  \country{Singapore}
  }
\email{ruijie_meng@u.nus.edu}

\author{Gregory J. Duck}
\authornote{Joint first author}
\affiliation{%
  \institution{National University of Singapore}
  \country{Singapore}
  }
\email{gregory@comp.nus.edu.sg}

\author{Abhik Roychoudhury}
\affiliation{%
  \institution{National University of Singapore}
  \country{Singapore}
  }
\email{abhik@comp.nus.edu.sg}

\begin{abstract}
Computer programs are not executed in isolation, but rather interact with the execution environment which drives the program behaviors.
Software validation methods thus need to capture the effect of possibly complex environmental interactions. Program environments may come from files, databases, configurations, network sockets, human-user interactions, and more. Conventional approaches for environment capture in symbolic execution and model checking employ environment modeling, which involves manual effort. In this paper, we take a different approach based on an extension of greybox fuzzing.
Given a program, we first record all observed environmental interactions at the kernel/user-mode boundary in the form of system calls.
Next, we replay the program under the original recorded interactions, but this time with selective mutations applied, in order to get the effect of different program environments---all without environment modeling. Via repeated (feedback-driven) mutations over a fuzzing campaign, we can search for program environments that induce crashing behaviors. Our \tool tool found 33 previously unknown bugs in well-known real-world protocol implementations and GUI applications. Many of these are security vulnerabilities and 16 CVEs were assigned.
\end{abstract}




\maketitle

\section{Introduction} \label{sec:introduction}

Computer programs are not executed in isolation, but rather interact with a complex {\em execution environment} which drives the program behaviors.
Inputs received from the environment, such as configuration files, terminal input, human-user interactions, and network sockets, directly affect the internal program state which, in turn, governs how the program executes.
Outputs sent to the environment, such as terminal output and sockets, provide useful clues that reflect these program states and behaviors.
If the program is buggy, some environmental interactions may cause the program to crash or otherwise misbehave.
{\em Fuzz testing} (or {\em fuzzing})~\cite{bohme2020fuzzing} is a widely-used automatic method that can find such program (mis)behavior.
Ideally, fuzzing should be executed under different execution environments to comprehensively explore diverse program behaviors.
However, capturing the effect of complex environments has always been a challenge for all program-checking methods---be it software verification, analysis, or testing. A dominant approach for handling different environments is {\em environment modeling}, which is used by verification and analysis methods including model checking and symbolic execution. 

Algorithmic verification methods, such as model checking, conduct a search over the space of program states. Thus to verify an open software system interacting with the environment, model-checking methods typically describe the environment as a separate process. This process captures an {\em over-approximation} of possible behaviors that could be exhibited by real concrete environments. The environment process is then composed with the open software system, forming a closed system which can then be subjected to search. Environment synthesis for model checking has been studied in works such as~\cite{env-mc}. These approaches depend on user-provided specifications to implement a {\em safe} approximation of the environment, and do not use concrete environments to demonstrate program errors.

In symbolic analysis methods \cite{baldoni2018survey, exe} and tools such as \klee \cite{klee}, environment capture is handled by redirecting calls to ``environment models''. These models are hand-written \verb+C+ code, specifically, the \klee paper \cite{klee} mentions writing 2500 lines of \verb+C+ code to implement 40 system calls. Note that even these are simplified descriptions of the system calls. Although this approach is modeling-based, these works show a more direct attempt to handle program inputs from different environmental sources such as files, networks, etc.

In this paper, we take a fresh look at the problem of program environment capture, and provide a solution in the context of fuzz testing. Greybox fuzzing uses a biased random search over the domain of program inputs to find crashes and hangs. We aim to extend greybox fuzzing over the {\em full environment} without resorting to modeling. Our approach is to first run the program normally, but also to {\em record} all interactions between the program and environment that can be observed at the user/kernel-mode boundary (e.g., {\em system calls}). These interactions serve as the set of initial seeds. Next, the program is iteratively run again as part of a fuzzing loop, but this time {\em replaying} the original recorded interactions. During the replay, the fuzzer will opportunistically mutate the interactions recorded for system calls to observe the effect of environments different from that of the original recording. In effect, the program environment is fuzzed at the system call layer. Our approach does not conduct any abstraction of possible environments; it  (implicitly) works in the space of real concrete environments. %, without storing them.

We present a generic approach for fuzzing the full program environment.
Existing greybox fuzzers are limited to fuzzing {\em specific} input sources, such as an input file specified by the command line (\eg~\afl~\cite{afl} and \aflplus \cite{afl++}), or a network socket over a specific network port (\eg~\aflnet \cite{aflnet} and \nyxnet \cite{nyx-net}).
Our approach extends the scope of fuzzing to include {\em all} environmental inputs, meaning that any input is considered a fuzz target, regardless of source.
We also propose a generic fuzzing algorithm to (implicitly) generate different program environments, thereby exploring diverse program behaviors.
We have implemented our approach of program environment fuzzing in the form of a new greybox fuzzer called \tool. 
We evaluate \tool against two categories of user-mode programs under Linux: network protocol implementations and GUI applications, both of which are considered challenging subjects for existing fuzzers \cite{afl++, bohme2020fuzzing}. In real-world and well-known applications, such as Vim and GNOME applications, \tool found 33 previously unknown bugs (24 bugs confirmed by developers, which include 16 new CVEs).
The bugs found include null-pointer dereferences,  buffer overreads, buffer overwrites, use-after-frees, and bad frees, all triggered by fuzzing diverse environmental inputs including sockets, configuration files, resources, cached data, etc.

In summary, we make the following main contributions:
\begin{itemize}[leftmargin=*]
    \item We propose a new greybox fuzzing methodology to capture the effect of complex program environments---all without environment modeling or manual effort. 
    \item We present a new fuzzing algorithm based on the full environmental record and replay at the user/kernel-mode boundary.
    \item We implemented the approach as a generic fuzzer (\tool) capable of testing various program types, including two categories of recognized challenging subjects. In our evaluation, we found 33 previously unknown bugs and received 16 CVE IDs.  
\end{itemize}
Our tool is publicly available at
\begin{center}
    \url{https://github.com/GJDuck/EnvFuzz}
\end{center}

\section{Background and Motivation} \label{sec:motivation}

\subsection{Motivating Example}
As an initial motivating example, we consider a calculator application implemented using a {\em Graphical User Interface} (GUI).
A human user makes inputs in the form of mouse movements, keystrokes, button presses, \etc, and the application reacts by generating outputs that update the graphical display.
For example, by pressing the button sequence $\langle\texttt{1},\texttt{+},\texttt{2},\texttt{=}\rangle$, the application responds by displaying the answer (\texttt{3}).

Like all software, the calculator application may contain bugs, and these bugs can be discovered using automatic software testing methods such as fuzzing.
For example, a fuzzer could apply the mutations $(\texttt{+}){\rightarrow}(\texttt{/})$ and $(\texttt{2}){\rightarrow}(\texttt{0})$ to construct a new button press sequence $\langle\texttt{1}, \texttt{/}, \texttt{0}, \texttt{=}\rangle$ that will cause a crash (\verb+SIGFPE+) if the calculator application were to not properly handle division by zero.
Although most mutations will be benign (non-crashing), typical fuzzers mitigate this with a combination of high throughput (\eg 100s of executions per second), program feedback (\eg code coverage), and power scheduling (\eg controlling mutation counts), increasing the likelihood of finding crashing inputs within a given time budget.

However, most existing greybox fuzzers, such as \afl~\cite{afl, afl++} and \aflnet~\cite{aflnet}, 
do not consider all input sources when producing mutated inputs. These fuzzers only target a specific class of inputs by default.
For example, \afl only targets standard input (\verb+stdin+) or a file specified by the command line.
Similarly, \aflnet only targets network traffic over a specific port for a specific popular protocol (\eg \texttt{ftp} and \texttt{smtp}).
Essentially, these existing fuzzers use a simplified program environment, where program behaviors (and potential crashes) are driven by a single input source, and it is up to the tool user to decide {\em which} input source to fuzz.
All other input sources are considered as ``static'', \ie unmutated and unchanged between test cases.
Furthermore, most existing fuzzers are specialized to specific {\em types} of inputs, such as regular files or popular network protocols.

\begin{figure}
    \centering
    \setlength{\belowcaptionskip}{-5pt}
    \includegraphics[page=1, scale=2.3]{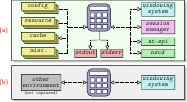}
    \caption{(a) is a calculator application with the {\em full} environment, including regular file I/O, standard streams, and socket/event fds to various system services.
    (b) is a {\em simplified} environment with a single input/output (windowing system socket), where all other interactions are not captured.}
    \label{fig:calc}
\end{figure}

In reality, most programs have a more complicated interaction with the environment beyond that of a single input source.
For example, if we consider the \verb+gnome-calcuator+ application as part of the GNOME Desktop Environment for Linux.
This application will open 706 distinct file descriptors under a minimal test (\ie open and close the application window), including:
\begin{itemize}[leftmargin=*]
\item[-] $674{\times}$ regular files, including
       configuration, cache,
       and GUI resources (icons/fonts/themes).
\item[-] $7{\times}$ socket connections to the windowing system, 
    session manager, and other services.
\item[-] Miscellaneous (\eg special files, devices, and \texttt{stderr}).
\end{itemize}
The calculator application with a full environment is illustrated in \autoref{fig:calc}~(a).

\subsection{Limitations of Conventional Fuzzing}

Fuzzing requires two key decisions to be made before use: 
\begin{itemize} [leftmargin=*]
\item \label{case:input}  \emph{Input Selection}:
   Which input should be fuzzed?
\item \label{case:env} \emph{Environment Modelling}:
   How to handle other inputs?
\end{itemize}
For the button-press example, the fuzz target would be the windowing system socket over which button-press events are received.
Thus, for the purposes of fuzzing, we use a simplified environment as illustrated in \autoref{fig:calc} (b).
In the case of the calculator application, the simplified environment is somewhat na\"ive, since the target socket is only one of many possible input sources (706 possibilities). Consequently, only a small fraction of the actual environment is subjected to fuzzing. Assuming, for the sake of example, that the windowing system socket is selected. The next step is to choose a fuzzer. Since the input is a socket rather than a file, a network protocol fuzzer, such as \aflnet, will be suitable. \aflnet works by fuzzing inbound network messages and parsing the response codes from outbound messages as feedback to guide the fuzzing process. However, \aflnet only supports a limited set of pre-defined network protocols, and this does not include support for windowing system protocols. Even if the necessary protocol support is available, the environment beyond the fuzz target must still be handled. 
% gjd: slight rewording to shorten

One approach is to fix all the remaining environments as most existing fuzzers do, where the program is consistently checked within a single environment across test cases. Obviously, this approach limits the explored program behaviors. Moreover, in some cases, such as fuzzing the calculator application and other GUI applications, this approach is impractical for existing fuzzers. Handling regular file I/O is relatively straightforward since files can be read from disk for each executed test case, with outputs easily discarded (e.g., piped to \verb+/dev/null+). However, a program can interact with more than one external service, such as session managers, service daemons, and even human users. In order to execute a single test case as part of the fuzzing process, the system-environment interactions would need to be ``reset'' for each individual test case---something known to be slow. For human-driven inputs, this also implies that a human-in-the-loop is necessary, since the fuzzer needs human interaction to proceed from one test execution to another.
% gjd: reworded the last sentence

Another approach is to build a {\em model} of possible environmental interactions. However, modeling is non-trivial. For example, each external service will typically use its own specialized protocol, and there can be an arbitrary number of services in the general case. Furthermore, any model would need to be {\em accurate}, as an invalid interaction may cause the test subject to terminate early due to an error condition, thus hindering reaching potential bug locations. Environment modeling is a known problem in the context of model checking and symbolic execution. Many existing works~\cite{ball2006thorough, klee} address it by modeling the environment {\em manually}. However, these approaches tend to be limited to specific problem domains and lack scalability for the general case.

\subsection{Core Idea} \label{sec:rrfuzzing}

We now describe our approach. We do not explicitly enumerate all possible environments in a search space and then navigate this very large search space. Our approach (below) is more implicit.
\begin{itemize}[leftmargin=*]
\item \emph{Input Selection}: All environmental inputs are fuzzed.
\item \emph{Environment Modelling}: Avoid modelling. The inputs are executed under a given environment and the effect of different environments is captured by mutating the environmental interactions represented by system calls. 
\end{itemize}
For the calculator example, we consider all environmental inputs as fuzz targets regardless of \emph{type}. Thus, various files (\eg configuration, cache, and resource), sockets (\eg those utilized by the windowing system), and any other input sources, are abstracted as generic {\em inputs} to subject fuzzing, eliminating the need for special handling. Since the whole environment is the fuzz target, any remaining residual environment is essentially eliminated, avoiding the need for additional modeling. 

Building upon this concept, our approach first {\em records} all environmental interactions between the target program and its environment. Subsequently, the program is iteratively run again as part of a fuzzing loop, this time by {\em replaying} the interactions from the previous recording to substitute the original environment. Instead of replaying exactly the original recording, some of the interactions are {\em mutated} to implicitly generate the effect of different program environments, potentially uncovering new program behaviors.  

To record the full environmental interactions, our approach works at the system call layer. This is motivated by the observation that most user-mode applications in Linux interact with the environment through the kernel/user-mode interface. For example, button presses and corresponding GUI updates flow through \texttt{recvmsg} and \texttt{sendmsg} systems calls over a socket. Similarly, pipes, streams and file I/O flow through standard \texttt{read} and \texttt{write} system calls. As a result, by recording system calls, we also record the full environmental interactions of the program, including system-environment interactions and human interactions, regardless of the underlying input type or the nature of the program. 

\section{System Overview} \label{sec:overview}

\begin{figure}[!t]
  \centering
  \includegraphics[page=1, trim= 0in 4.1in 0.1in 0in, clip, scale=0.65]{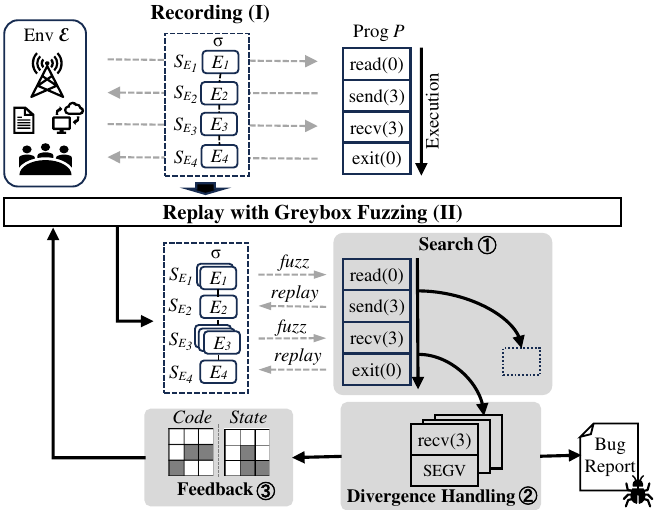}
  \caption{Overview of Program Environment Fuzzer \toolBold.}
  \label{fig:overview}
  \vspace{-10pt}
\end{figure}

Based on our core idea, we design a generic program-environment fuzzer using a {\em record and replay} methodology. This fuzzer, called \tool, is illustrated in \autoref{fig:overview}. At a high level, \tool consists of two phases: the \emph{record} phase (Phase~\RN{1}) records the full interactions between the program-under-test $P$ and its environment $\mathcal{E}$, and the \emph{replay-with-greybox-fuzzing} phase (Phase~\RN{2}) replays and fuzzes the recorded interactions. Phase~\RN{2}
captures the effect of different program environments and uncovers new program behaviors.

\subsubsection*{\textbf{Phase~\RN{1}: Recording}}

In the \emph{recording} phase, the program $P$ is run normally within some test environment $\mathcal{E}$. The program interacts with the environment (files, sockets, human input, etc.) via a sequence of system calls, which are intercepted by \tool and saved into a {\em recording} $\sigma$. Here, $\sigma$ is an in-order sequence of {\em records} (\eg $\sigma$ = [$E_1$, $E_2$, $E_3$, $E_4$]), where each record $E$ stores all of the necessary details for reconstructing each corresponding environmental interaction in Phase~\RN{2}. These details include the system call number, system call arguments, buffer contents (if applicable), and the return value. These records are then saved into a respective seed corpus ($\mathcal{S}_E$ = \{$E$\}) to serve as the initial seeds for the subsequent replay and greybox fuzzing. 

\subsubsection*{\textbf{Phase~\RN{2}: Replay with Greybox Fuzzing}}

Phase~\RN{2} combines environment replay with greybox fuzzing. 
The idea is two-fold:
\begin{enumerate}[leftmargin=*]
\item Faithfully \emph{replay} the recorded environment interactions to reconstruct (deep) program states observed during Phase~\RN{1};
\item \emph{Fuzz} each reconstructed state using greybox fuzzing.
\end{enumerate}
Faithful replay works by re-running the program, but using the recording $\sigma$ as a substitute for the original test environment $\mathcal{E}$.
This again works by intercepting system calls, but this time the corresponding record ($E \in \sigma$) is {\em replayed} as a substitute for the real interaction.

To uncover different program behaviors for bug discovery, the core of \tool lies in greybox fuzzing. However, unlike traditional fuzzers, \tool works by fuzzing the recorded environmental interactions ($E \in \sigma$), rather than targeting specific files, as with \afl, or sockets, as with \aflnet. This works as follows: for each state reconstructed, \tool faithfully replays the next environmental interaction $E$ in sequence from $\sigma$ to advance state reconstruction. \emph{In addition}, \tool selects seeds from the seed corpus $\mathcal{S}_{E}$, assigns energy, and introduces mutations to generate {\em mutant} interactions. Each mutant interaction is replayed in a {\em forked} branch of execution, where the program's behavior is observed (see \autoref{fig:overview}~\ccc{1}). 

Following the execution of a mutant interaction, the program behavior may {\em diverge} significantly from the original recording.
Such divergence can include the program invoking different system calls, or invoking existing system calls but in a different order.
For example, as shown in \autoref{fig:overview}, the \verb+exit(0)+ system call could be changed into \verb+recv(3)+.
Such behavior divergence presents a technical challenge for advancing replay, since only the original recording ($\sigma$) is available.
Indeed, the main goal of fuzzing is to explore novel (divergent) program behaviors in order to discover bugs.
To resolve this challenge, \tool introduces the notion of \emph{relaxed} replay (as opposed to \emph{faithful} replay) that is designed to progress divergent program execution after mutation (see \autoref{fig:overview}~\ccc{2}).

At the end of each execution, similar to traditional greybox fuzzing, program feedback is used to determine {\em interesting} mutant interactions (see \autoref{fig:overview}~\ccc{3}). The interesting interactions are saved into the seed corpus for future mutation. Additionally, mutations triggering program crashes are saved and reported to the user. The fuzzing campaign repeatedly iterates over reconstructed program states until a time budget is reached.

\section{Environment Fuzzing}\label{sec:approach}

We describe and explain the \tool algorithm in this section.

\subsection{Environment Recording and Replay} \label{sec:record}

For recording the environment, \tool implements a system call {\em interceptor routine} that acts as a proxy (i.e., ``man-in-the-middle'') between the program $P$ and the kernel.
Thus, when the program invokes a system call, such as a \verb+read+ or \verb+write+, the call will be routed to the interceptor routine.
The routine first \emph{forwards} the system call to the underlying kernel and waits for the result.
Once the underlying system call completes, the interceptor routine will then save relevant information about the system call into a record $E$, including: the system call number (e.g., \verb+read+ and \verb+write+), arguments (e.g., file descriptor, buffer pointer, and buffer size), buffer contents (where applicable), current thread ID, and the return value.
The system call result is then returned back to the program $P$, which continues executing as normal.

Each individual record $E$ represents an interaction between the program $P$ and its environment $\mathcal{E}$.
During recording, each record is appended onto an in-order sequence $\sigma$, otherwise known as the \emph{recording}, and is also saved into the respective seed corpora.
The recording $\sigma$ contains the information necessary to reconstruct all program states previously observed during the recording phase.
For \emph{faithful} replay, the program is run once more, but this time
the interceptor routine instead \emph{replays} (rather than forwards) the previously-recorded $E$.
For fuzzing, the original record is replayed, but with one or more mutations applied first.
Such mutations represent modified environmental interactions, and can change the program behavior.

We now use an example to illustrate this process. Suppose that during recording, the program $P$ calls $\texttt{read}(0,\mathit{buf},100)$, which is forwarded to the kernel, and the user enters ``\texttt{quit{\textbackslash}n}'' into \verb+stdin+ ($\mathit{fd}{=}0$).
The interceptor routine will record the returned buffer contents (``\texttt{quit{\textbackslash}n}'') and the returned value (${=}5$ bytes read) into a record $E$. Then, during replay with greybox fuzzing:
\begin{itemize}[leftmargin=*]
\item
For \emph{faithful} replay, the program $P$ is re-run, and calls the same $\texttt{read}$ system call as before.
Instead of forwarding the system call to the kernel, the interceptor routine copies the previously recorded contents from $E$, copying ``\texttt{quit{\textbackslash}n}'' into $\mathit{buf}$ and returning $5$.
This causes the program's execution to proceed equivalently to the original recording.
\item
For fuzzing, the record $E$ is first {\em mutated} before it is replayed.
For example, the buffer contents could be mutated into ``\texttt{quip{\textbackslash}n}'', and this will likely cause the program's behavior to diverge as if this were the original user interaction---possibly exposing new behaviors and bugs.
\end{itemize}
The mutation is applied to the buffer contents of {\em input} system calls (\eg \verb+read+) as this can affect the program behaviors and cause behavior divergence.
Other system calls, that do not affect the program behaviors (\eg \verb+write+), will not be mutated.
The combination of faithful replay, and replay with mutation, forms the basis of \tool's greybox fuzzing algorithm.

\begin{figure*}
    \centering
\begin{minipage}{0.33\textwidth}
\footnotesize
\centering
\begin{lstlisting}[frame=single]
  int main()   {
     char buf[SIZE]; int r;
     r = read( 0, buf, SIZE); ...;
     r = send( 3, buf, r);    ...;
     r = recv( 3, buf, SIZE); ...;
     r = send( 3, buf, r);    ...;
     r = recv( 3, buf, SIZE); ...;
     r = write(1, buf, r);    ...;
     exit(0); }
\end{lstlisting}
\end{minipage}
%\hfill
\hspace{0.2cm}
\begin{minipage}{0.6\textwidth}
    \includegraphics[scale=2.58]{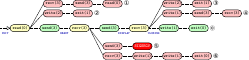}
\end{minipage}
    \caption{Illustration of the underlying fuzzing algorithm.
    Here, the example program reads from file descriptor \texttt{0}, then interacts with socket (file descriptor \texttt{3}).
    The fuzzer faithfully \emph{replays} a previously \emph{record}ed interaction \ccc{0}, as well as several \emph{mutant} interactions \ccc{1}/\ccc{2}/\ccc{3}/\ccc{4}/\ccc{5}/\ccc{6}.
    Each mutant interaction is generated by mutating at least one input system call from the faithful replay.
    This causes the program's behavior to diverge, including exit with error \ccc{2}/\ccc{3}, system call reordering \ccc{1}/\ccc{6}, new I/O system call \ccc{4}, and a crash \ccc{5}.
    The program state $\{\texttt{INIT},\texttt{READY},\texttt{DISPLAY},\texttt{CLOSING}\}$ between select system calls is also illustrated.
    \label{fig:example}}
\vspace{-5pt}
\end{figure*}

\subsection{Reflections on Search Challenges}

After the recording phase, \tool has collected a set of initial seeds representing real environment interactions.
Using these as a basis, \tool employs greybox fuzzing to generate new interesting seeds representing interactions with new program environments---each with the potential to induce novel program behaviors. In designing an efficient algorithm for searching the program environment space, there are two main challenges:
(i)~\emph{statefulness}: how to effectively explore deep program behaviors? 
(ii)~\emph{throughput}: how to maintain high fuzzing throughput?

\subsubsection*{\textbf{Challenge (i): Statefulness}} To better understand Challenge (i), we consider a calculator program that interacts over multiple input and output sources, as shown in \autoref{fig:example}. The program begins in an \verb+INIT+ state, where it first parses a configuration file (file descriptor \texttt{0}), then creates a user interface (GUI) by sending a message over the windowing system socket (file descriptor \texttt{3}).
The program then transitions into a \verb+READY+ state---i.e., waiting to accept mathematical expressions from the user interface. Subsequently, the program processes one user input expression (received from \texttt{3}), and then sends the result back to the interface (send to \texttt{3}), and the program transitions into the \verb+DISPLAY+ state. Finally, the user closes the interface (received from \texttt{3}), and the program transitions into a \verb+CLOSING+ state. Here, the program writes a message to the terminal (file descriptor \texttt{1}) before exiting.
Our example is a simplification for brevity, as a real calculator program will typically interact with thousands of system calls, and may have many more internal states.

At the layer of system calls, the program is {\em stateful}, as it accepts a sequence of environmental inputs and adjusts its state accordingly. Some program behaviors are only reachable by specific states, which are in turn reachable only through specific input sequences. When fuzzing stateful programs, greybox fuzzers aim to exercise each observed state in order to explore the neighborhood of potential program behaviors, thereby having a greater chance to expose new bugs. However, state identification remains a challenging problem for fuzz testing in general. Existing works ~\cite{ijon, aflnet, nsfuzz, stateful} propose several heuristics for program state detection. For example, \ijon requires states to be manually annotated, whereas \aflnet utilizes response codes from outbound messages to detect new states for well-known protocols. Either way, existing approaches require manual effort or are specialized to specific input sources.

We propose a generic approach that considers all input sources, such as files, sockets, and pipes, and consider how they affect program states. We consider each input system call as a \emph{potential} state transition. For example, in \autoref{fig:example}, after executing each input system call in sequence, the program transitions from the \verb+INIT+ to \verb+READY+ state, then from the \verb+READY+ to \verb+DISPLAY+ state, and finally from the \verb+DISPLAY+ to \verb+CLOSING+ state.
Thus, each input can be fuzzed as a distinct transition between states, regardless of the input type (file, socket, etc.).
However, some of the inputs may not trigger new transitions. This is mitigated by the power schedule~\cite{aflfast}, where inputs that fail to induce real state transitions are also less likely to expose new program behaviors observable via program {\em feedback}.
As such, the corresponding input will be assigned less {\em energy}, and is naturally deprioritized for future mutations. 

\subsubsection*{\textbf{Challenge (ii): Throughput}} To reach each observed state, \tool conducts a faithful replay of the recorded system calls. Upon reaching a state (\ie before executing the corresponding input system call), \tool applies mutations to explore the neighboring program behaviors.
If the fuzzer must always replay system calls from the root point, then multiple system calls need to be replayed to reach a specific state.
For example, in the calculator example, a total of four system calls must be faithfully replayed to reach the \verb+DISPLAY+ state.
This can significantly slow fuzzing throughput, especially for real-world examples where thousands of system calls may be required to reach a given state.
To address this challenge, we propose a tree-based fuzzing algorithm that avoids (re)executing the same prefix sequence of system calls repeatedly. 
The algorithm is illustrated by the tree shown in \autoref{fig:example}.
Specifically, the original recording is faithfully replayed (without mutation), forming the ``spine'' of the tree, which is represented by the middle trace \ccc{0}. Upon reaching an input system call, \tool additionally \verb+fork+s-off some number of {\em mutant} traces, creating the ``branches'' of the tree (\eg \ccc{1}/\ccc{2}). Each branch starts by replaying an original input with one or more {\em mutation} operators applied, and may involve further mutations of subsequent inputs. After executing each branch, \tool continues the faithful replay to grow the spine until the next input point, after which \tool \verb+fork+s-off more branches (\eg \ccc{5}/\ccc{6}).
The process repeats once more (\eg \ccc{3}/\ccc{4}) before \ccc{0} terminates.

\begin{algorithm}[t]
    \caption{Program Environment Fuzzing Algorithm.\label{alg:overview}}
    \DontPrintSemicolon
    \small
    \newcommand\mycommfont[1]{\ttfamily\textcolor{blue}{#1}}
    \SetCommentSty{mycommfont}
    \SetNoFillComment
    \SetKwProg{Fn}{func}{:}{}
    \SetKwInOut{Input}{Input}
    \SetKwInOut{Output}{Output}
    \SetKwInOut{Globals}{Globals}
    \SetKwFunction{FuzzEvent}{FuzzEvent}
    \SetKwFunction{Instrument}{Instrument}
    \SetKwFunction{RRecord}{Record}
    \SetKwFunction{RReplay}{FuzzReplay}
    \SetKwFunction{EFuzz}{EnvFuzz}
    \SetKwFunction{Syscall}{syscall}
    \SetKwFunction{SyscallFuzz}{FuzzSyscall}
    \SetKwFunction{Replay}{ReplaySyscall}
    \SetKwFunction{Emulate}{EmulateSyscall}
    \SetKwFunction{IsInteresting}{IsInteresting}
    \Input{Program $P$, environment interaction $\mathcal{E}$}
    \Output{Crashing events \textit{\crashingSeeds}}
    \Globals{Input-specific corpora $\mathcal{S}_E$}
    \vspace{0.5em}
    \Fn(\label{line:rrbegin}){\EFuzz{$P$, $\mathcal{E}$}}{
    $\sigma \gets~$\RRecord{$P$, $\mathcal{E}$} \MyComment{\footnotesize{Recording}}\label{line:rrecord} \\
    \lFor{$E \in \sigma$}{$\mathcal{S}_E \gets \{E\}$} \label{line:rrecord-corpus}
     \Repeat{\textup{timeout reached or abort}\label{line:outend}}{\RReplay{$P$, $\sigma$}\label{line:rreplay}}
    }
    \Fn(\MyComment{\footnotesize{Replay with Fuzzing}}){\RReplay{$P$, $\sigma$}}{
        $exec(P_{[\text{replace \Syscall with \SyscallFuzz}, \mathit{isBranch}{\leftarrow}\mathit{false}]}, \sigma)$\label{line:exec}
    }\label{line:rrend}
    \vspace{0.5em}
    \Fn(\label{line:fuzzbegin}\MyComment{\footnotesize{Tree-based Search}}){\SyscallFuzz{$e$}}{
            \uIf{$\mathit{isBranch}$}{
                \Return \Emulate{$e$, $\sigma$}\label{line:emulate} \MyComment{\footnotesize{Divergence Handling}}
            }
            \Else(\tcc*[h]{\textbf{\textrm{if}}\hspace{0.2em}$\mathit{isSpine}$\hspace{0.3em}\textbf{\textrm{then}}}\label{line:spinebegin}){
             $E \gets \mathit{head}\textup{(}\sigma\textup{)}$; $\sigma \leftarrow \mathit{tail}\textup{(}\sigma\textup{)}$ \label{line:retrive-seed}\\  
             \lIf{$\neg \mathit{isInput}$\textup{(}$e$\textup{)}}{
                \hspace{-0.3em}\Return \Replay{$E$} \label{line:notinput}\label{line:spine1}
             }
             \For{$E' \in \mathcal{S}_E, i \in 1..\mathit{energy}(E')$ \label{line:power-scheduling}}{
                  $E'' \leftarrow \mathit{mutate}(E')$\label{line:mutate}\; 
                  $\mathit{pid} \gets \texttt{fork}\textup{()}$\label{line:fork}\;
                  \uIf(\MyComment{\footnotesize{In child:}}){$\mathit{pid} = 0$} 
                  {
                      $\mathit{isBranch} \gets \mathit{true}$ \\
                      \Return \Replay{$E''$}\label{line:branch}
                  }
                  \Else(\MyComment{\footnotesize{In parent:}}){
                      $\texttt{waitpid}\textup{(}\mathit{pid}\textup{,}\, \texttt{\&}\mathit{status}\textup{)}$\label{line:await}\; 
 %                     $res \gets \mathit{processResult}\textup{(}\mathit{status}\textup{)}$\; 
                      \lIf{$\mathit{isCrash}\textup{(}\mathit{status}\textup{)}$}{
                            \hspace{0.425em}add $E''$ to \crashingSeeds \label{line:crash}
                      }
                      \lIf(\label{line:inend}){$\mathit{isInteresting}\textup{(}E''\textup{)}$}{
                            \hspace{-0.25em}add $E''$ to $\mathcal{S}_E$ \label{line:interesting}
                      }
                  }
               }
               \Return \Replay{$E$} \MyComment{\footnotesize{Grow Spine}}\label{line:spineend}\label{line:spine2}
            }
        }\label{line:fuzzend}
\end{algorithm}

\subsection{Fuzzing Search Algorithm} \label{sec:tree_search}

Based on the environment recording and replay technique, along with the efficient search strategy, we introduce a novel environment fuzzing algorithm, illustrated in \autoref{alg:overview}.
The recording is shown in \autoref{line:rrecord} of \autoref{alg:overview}.
After the recording, the program is executed normally, but with the interceptor routine $\mathtt{FuzzSyscall}$ replacing the standard system call interface (\autoref{line:exec}).
There are two main cases to consider: the replay is in the spine or in a branch (\eg see \autoref{fig:example}), and the program starts with running in the spine ($\mathit{isBranch}{\leftarrow}\mathit{false}$). 
For the spine of the tree (\autoref{line:spinebegin}-\autoref{line:fuzzend}), \tool retrieves the next record $E$ to be processed (\autoref{line:retrive-seed}). For non-input system calls (\eg \verb+write+), the original record $E$ is faithfully replayed ``as-is'' (\autoref{line:notinput}).
Conversely, all input system calls (\eg \verb+read+) are treated as potential fuzzing targets, and a greybox fuzzing algorithm is used (\autoref{line:power-scheduling}-\autoref{line:interesting}).
Specifically, for each record $E$ corresponding to the input syscall, \tool will iterate over each seed $E'$ from corpus $\mathcal{S}_E$. % for mutation.
For each $E'$, \tool applies one or more standard {\em mutation operators}, to further mutate the input buffer contents, and thereby generating a new seed $E''$ (\autoref{line:mutate}). The current implementation uses mutation operators from other fuzzers, \eg \verb+havoc+ from \afl \cite{afl, aflnet}. The number of mutations is controlled by a power schedule ($\mathit{energy}$) (\autoref{line:power-scheduling}). 

To execute the new seed $E''$, the algorithm first forks the program into a {\em parent} and {\em child} process (\autoref{line:fork}). The seed $E''$ is executed in the child, forming a {\em branch} of the tree (\eg see \autoref{fig:example}), while the parent waits for the child's termination (\autoref{line:await}). 
After applying a mutation in the child, the interceptor routine $\mathtt{FuzzSyscall}$ processes the subsequent system calls using a different method (\autoref{line:emulate}), which will be discussed in \autoref{sec:divergence_handling}.
Following the termination of the child, the parent examines the result. Crashing mutations are saved into a special corpus \crashingSeeds that forms the output of \autoref{alg:overview} (\autoref{line:crash}). Otherwise, the fuzzing feedback (discussed in \autoref{sec:feedback}) is used to determine whether the mutated seeds are {\em interesting} or not, and interesting seeds are saved into $\mathcal{S}_E$ for future mutation see \autoref{line:interesting} ; the decision on whether a seed is interesting or not, is conducted based on fuzzing feedback which is discussed in the next subsection.  Subsequently, \tool grows the spine by continuing faithful replay (\autoref{line:spine2}). 
After the fuzzing campaign is complete, the \tool infrastructure also supports replaying any of the \crashingSeeds corpus to reproduce discovered bugs.

An illustration of this fuzzing algorithm on a simple example program appeared in \autoref{fig:example}.

\subsection{Fuzzing Feedback} \label{sec:feedback}

Greybox fuzzing relies on feedback to select ``interesting'' seeds (\autoref{line:interesting} in \autoref{alg:overview}) to guide the search towards novel program behaviors, thereby increasing the likelihood of discovering bugs \cite{evaluating}. A common form of feedback is {\em branch coverage}, as used by many modern fuzzers~\cite{afl, afl++}. Here, seeds that cover new branches (code paths) have the potential to explore different behaviors, and thus are considered interesting and saved into the corpus for future mutation. 
Most fuzzers collect branch coverage feedback using compiler instrumentation (\eg \verb+afl-gcc+).
Instrumentation can also be inserted directly into binary code using {\em static binary rewriting}, such as with \eafl~\cite{e9afl}. \tool supports branch coverage feedback and operates directly on binaries to maximize generality.

For the case of stateful programs, branch coverage alone is generally considered insufficient~\cite{ijon, aflnet}.
As such, {\em state feedback} has been proposed in collaboration with branch coverage to guide the fuzzing process.
Here, seeds that cover new state transitions are also considered ``interesting'' and are similarly added to the corpus.
However, as discussed in \autoref{sec:tree_search}, automatically inferring program states is challenging, especially for binary code.
Our approach is to treat each input message as a {\em potential} state transition. 
We leverage program {\em outputs} (\eg \verb+write+)  as a proxy for detecting states. Our heuristic is that, under certain inputs, a program will generate output that is contingent on its internal states, and thus outputs can provide insights into these states. To mitigate the impact of outputs with unknown structures/formats, we employ locality-sensitive hashing and clustering based on the {\em Hamming distance} \cite{stateafl, snipuzz}.
\tool can utilize both branch and state feedback to guide the search.

\section{Relaxed Replay for Divergence}\label{sec:divergence_handling}

After a mutated input is replayed in a branch, it is common for the program's behavior to {\em diverge} from the original recording, as illustrated by the branches \ccc{1},...,\ccc{6} in \autoref{fig:example}. Divergence could include: exiting with error \ccc{2}/\ccc{3}, system call reordering \ccc{1}/\ccc{6}, new system call invoking \ccc{4}, or even the program crashing \ccc{5}. For example, suppose the last input from \autoref{fig:example} receives a command ``\texttt{quit{\textbackslash}n}'' from the socket, causing the program to enter the \verb+CLOSING+ state and exit. However, mutant replay could change the command to ``\texttt{quip{\textbackslash}n}'', foiling the state transition, and causing the program's behavior to diverge from the original recording.

This poses a challenge that is described as follows. During the recording phase, \tool will construct an in-order sequence of records $\sigma$. Assuming that $\sigma = [\sigma_1, E, \sigma_2]$, where $E$ is an input, then during the fuzzing phase, \tool faithfully replays the prefix $\sigma_1$ (as part of the spine) before reaching $E$. Next, \tool mutates $E$ to generate one (or more) mutant $E'$, after which $E'$ is replayed as a substitute for $E$. After replaying $E'$, the faithful replay of $\sigma_2$ may no longer be possible due to program behavior divergence, \ie the mutant sequence $\sigma' = [\sigma_1, E', \sigma_2]$ may be {\em infeasible}. The problem is that \tool only has the original recording $\sigma$ to work with.

To address this problem, we introduce the notion of {\em relaxed replay}. The key idea is to use system call {\em emulation} (\autoref{line:emulate} in \autoref{alg:overview}) to construct continuations of program execution that diverge from $\sigma$. Relaxed replay uses a set of {\em emulation routines}, one for each syscall number, where each routine takes the syscall arguments and returns a result (\ie return value, buffer contents) on a ``best-effort'' using available information. Unlike faithful replay, these routines can be called at any time and in any order, and do not necessarily need to follow the original recorded system call ordering. Crucially, the emulation routines should only return {\em plausible} results---\ie there exists a real (modified) environment $\mathcal{E}'$ from which the result could occur. Plausibility is necessary to avoid {\em false positives}---\ie crash reports that are irreproducible under any real environment. For plausibility, we generalize assumptions used by existing fuzzers, wherein any I/O modification (\eg in \afl) and reordering (\eg in \aflnet) are considered plausible. We now discuss these two cases in detail.

\begin{figure}
    \centering
    \includegraphics[scale=3.2]{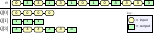}
    \caption{Illustration of the {\em global} ordering ($\sigma$) for faithful replay and a {\em local} ordering ($Q$) for relaxed replay.
    The relaxed replay partitions $\sigma$ into a set of miniqueues ($Q[\mathit{fd}]$) indexed by the file descriptor, each of which defines a local ordering specific to each $\mathit{fd}$.}
    \label{fig:desync}
\end{figure}

\subsection{Relaxing I/O System Call Ordering}

After mutation, programs often invoke I/O system calls in a different order from that of the original recording. To handle this case, our approach is to first {\em partition} $\sigma$ into a set of \emph{miniqueues} $Q[\mathit{fd}]$, with one miniqueue specific to each I/O source (\ie~{\em file descriptor}, $\mathit{fd}$).
The approach is illustrated by example in \autoref{fig:desync}. Here, under the \emph{global ordering} ($\sigma$) for faithful replay, only a read system call from file descriptor 0 can be serviced. However, after mutation, the program may attempt I/O on a different file descriptor. To handle such cases, our approach allows I/O system calls to be directly serviced from the corresponding miniqueue $Q[\mathit{fd}]$ under a {\em local ordering} specific to each $\mathit{fd}$, rather than the original global ordering ($\sigma$). The partitioning and local ordering is plausible under the assumption that I/O system calls can be reordered.

It is also common for programs to use the \verb+poll+ system call\footnote{See the \texttt{poll} manpage for more information.} to query which I/O operations are currently possible. Relaxed replay must also handle the poll system call using emulation. The algorithm is shown in \autoref{alg:poll}, and is a concrete example of an emulation routine. Here, poll is emulated based on the {\em current} state of $Q$ (\autoref{line:pollbegin}-\autoref{line:pollend}) and returning:

\noindent
(i) {\em End-of-file} (\verb+POLLHUP+) for an empty miniqueue (\autoref{line:pollhup});\\
(ii) {\em input ready} (\verb+POLLIN+) or {\em output ready} (\verb+POLLOUT+) if the queue head matches the requested event
         (\autoref{line:poll1}-\autoref{line:poll2});\\
(iii) \verb+0x0+ (a.k.a. no event) otherwise.

\noindent
If at least one of the returned events is non-zero, then the poll operation successfully completes (\autoref{line:done}) and execution continues. Otherwise, the poll operation will block. To avoid blocking, the algorithm heuristically {\em picks} a file descriptor and {\em reorders} the corresponding miniqueue (\autoref{line:reorder}), allowing \autoref{alg:poll} to always terminate (without blocking) in the next iteration of the outer-loop.

\subsection{Relaxing I/O System Calls}

Input system calls are emulated by an implicit \verb+poll+ operation, followed by popping the corresponding miniqueue $Q[\mathit{fd}]$. The popped record is replayed, possibly subject to further mutation. If the implicit poll operation indicates the miniqueue is empty (\verb+POLLHUP+), the input system call returns 0 indicating an end-of-file (\verb+EOF+).

Emulated output system calls similarly pop the corresponding miniqueue, but always succeed even if the queue is empty. This handles the common case where a mutation causes the program to generate additional output, such as a warning or error message that is not present in the original recording $\sigma$. Modified or extraneous outputs can generally be ignored, as outputs do not affect the program behavior. However, outputs do provide useful hints about the program state, which is used as fuzzing feedback. 

\begin{algorithm}[t]
    \caption{Emulated \texttt{poll} routine.\label{alg:poll}}
    \DontPrintSemicolon
    \small
    \newcommand\mycommfont[1]{\ttfamily\textcolor{blue}{#1}}
    \SetCommentSty{mycommfont}
    \SetNoFillComment
    \SetKwProg{Fn}{func}{:}{}
    \SetKwInOut{Input}{Input}
    \SetKwInOut{Output}{Output}
    \SetKwInOut{Globals}{Globals}
    \SetKwFunction{Poll}{EmulatePoll}
    \Input{Array of \texttt{pollfd} structs, $Q$ derived from $\sigma$}
    \Output{Number of non-zero $\mathit{revents}$}
    \vspace{0.5em}
    \Fn{\Poll{$\mathit{fds}$, $Q$}}{
      \While{$\mathit{true}$}
  {
  $r \leftarrow 0$; $h \leftarrow 0$ \label{line:pollbegin}\\
  \For{$i \in 0..|\mathit{fds}|{-}1$}{
      $E \leftarrow \mathit{head}(Q[\mathit{fds}[i].\mathit{fd}])$ \\
      % $Q \leftarrow \mathit{getMiniQueue}(\mathit{fds}[i].\mathit{fd}, \sigma)$ \\
      \uIf{$E=EOF$}{$\mathit{fds}[i].\mathit{revents}~\texttt{=}~\texttt{POLLHUP}$; $h\texttt{++}$ \label{line:pollhup}}
      \Else{
%        $E \leftarrow \mathit{head}(Q[\mathit{fd}])$ \\
%        $\mathit{fds}[i].\mathit{revents}~\texttt{=}~\mathit{fds}[i].\mathit{events}$ \\
        $\mathit{fds}[i].\mathit{revents}~\texttt{=}~\mathit{fds}[i].\mathit{events}~\texttt{\&}$\label{line:poll1} \\ 
        $\qquad(\mathit{isInput}(E)\texttt{?}~\texttt{POLLIN:}~\texttt{POLLOUT})$ \label{line:poll2} \\
        $r~\texttt{+=}~(\mathit{fds}[i].\mathit{revents}\texttt{?}~1\texttt{:}~0)$
      }
  }
  \lIf{$r > 0 \vee h > 0$}{\Return $r$} \label{line:done}\label{line:pollend}
  $\mathit{fd} \leftarrow \mathit{pick}(\mathit{fds},Q)$; $Q[\mathit{fd}] \leftarrow \mathit{reorder}(Q[\mathit{fd}])$ \label{line:reorder} 
  }
  }
\end{algorithm}

\subsection{Relaxing Non-I/O System Calls}

Other system calls are handled using heuristics, such as:
\begin{itemize}[leftmargin=*]
\item[-] \emph{Emulate}: emulate (plausible) effects of the system call;
\item[-] \emph{Forward}: pass the system call ``as-is'' to the underlying O/S;
\item[-] \emph{Fail}: fail the system call with an error condition (\eg \verb+ENOSYS+).
\item[-] \emph{Exit}: as a last resort, terminate the branch with \verb+exit+.
\end{itemize}
In addition to I/O system calls, \tool also implements several specialized emulation routines for other common system calls, including time (\eg \texttt{clock\_gettime}/etc.) and thread-related (\eg \texttt{futex}/\texttt{clone}/etc.) system calls.
For example, time-related system calls are handled by emulating a global {\em monotonically increasing} clock $t$.
The clock $t$ is first initialized to the last time observed in the recording before the branch, and $t$ is then incremented for each subsequent emulated system call after the branch.
This ensures that emulated time-related system calls always return \emph{plausible} results---\ie the system time always flows forwards.
The system calls related to memory management (\eg \texttt{brk}/\texttt{mprotect}/\texttt{madvise}/etc.) are generally \emph{forwarded} to the underlying O/S ``as-is'' without special handling.

Sometimes neither system call emulation nor forwarding is applicable.
For example, due to behavior divergence, the program may attempt to access a file that does not appear in the original recording ($\sigma$).
As such, \tool has no information about the file contents, or whether the file even exists.
In such cases,
relaxed replay can \emph{fail} the system call with an error (\eg \texttt{ENOSYS}), allowing for execution to proceed and giving the program a chance to recover. 
As a last resort, relaxed replay will exit the program if no other alternative is possible.
This occurs when a program ignores failure, e.g., by re-invoking the same system call again in a loop.

\section{Implementation}

We implemented the approach of \tool as a generic program environment fuzzer that can handle a diverse range of user-mode Linux applications, including GUI applications and network servers.
\tool is built on top of a full environment record and replay infrastructure, similar to that of \texttt{rr}-debug~\cite{rrdebug}.
In total, the \tool toolchain is implemented in over $\sim$13k source lines of \verb_C++_ code. 

The recording phase records all information that is necessary to faithfully replay the program $P$ during fuzzing.
In addition to system calls (the main focus of our discussion), the recording also includes additional information, such as the command-line arguments, environment variables, signals, thread interleavings, and special non-deterministic instructions (e.g., \texttt{rdtsc}).
System call interception is implemented using a variety of techniques.
The common case is handled using {\em static binary rewriting} to rewrite the \verb+syscall+ instruction in \verb+libc+, which diverts control-flow to the framework's interceptor routine.
For this, we use the \epatch~\cite{e9patch} binary rewriting system.
In addition, the framework also rewrites the \emph{virtual Dynamic Shared Object} (vDSO) at runtime, and also uses \verb+seccomp+ to generate a signal (\verb+SIGSYS+) that is used to intercept system calls outside of \verb+libc+ (less common case).
Our framework does not use \texttt{ptrace}, and thus avoids kernel/user-modes switches during replay for the common case.

Multi-threaded programs are handled by \emph{serializing} system calls during the recording phase, meaning that only one thread will run at a given time.
The recording phase runs the program normally using serialized system threads, whereas the replay-with-fuzzing phase uses lightweight \emph{fibers} as a replacement of system threads.
This design avoids one of the main technical limitations of \verb+fork()+, namely, that only the callee system thread will actually be cloned during a fork operation.\footnote{See the \texttt{fork} manpage for more information.}
In contrast, fibers are threads of execution that are implemented purely in user-mode, and where context switching is determined by the recorded schedule ($\sigma$).
Since there is no user-kernel interaction during replay, fibers can be used as a drop-in replacement of system threads without special handling.
Furthermore, since fibers are purely implemented in user-mode, they survive the fork operation intact, which is necessary for the \tool fuzzing algorithm.

\tool is also designed to operate directly on binaries without the need for source code. \tool uses both state and branch coverage as feedback. To collect the branch feedback, binaries can be instrumented using a modified version of \eafl~\cite{e9afl}. State coverage feedback does not require instrumentation.
Our implementation can record and fuzz large applications, including the subjects listed in our evaluation.

\section{Evaluation} \label{sec:evaluation}

To evaluate the effectiveness of \tool, we seek to answer the following research questions:

\begin{description}
\item [\textbf{RQ.1}] \textbf{New bugs.} Can \tool find previously unknown bugs in real-world and widely-used programs? Is fuzzing the program environment necessary to reveal these bugs?
\item [\textbf{RQ.2}] \textbf{Comparisons.} How many additional bugs does \tool discover over the baseline? How much more code coverage does \tool achieve compared to the baseline? Are the additional bugs and code coverage improvements related to program environment fuzzing? How many tests can \tool execute per second compared to the baseline?
\item [\textbf{RQ.3}] \textbf{Ablations.} What is the impact of each component on the performance of \tool?
\end{description}

\subsection{Experiment Setup} \label{sec:exp_setup}

\subsubsection*{\textbf{Subject Programs}} \label{sec:benchmark}

\tool is a generic fuzzer capable of testing a broad spectrum of user-mode programs in Linux.
Given the scope of applications that \tool can fuzz, we shall focus on two core categories of program: network protocols and \emph{(Graphical) User Interface} GUI/UI applications that interact with a human user via the windowing system or terminal.
These two categories have been recognized as challenging for fuzzing~\cite{bohme2020fuzzing}.
For example, fuzzing GUI applications with \aflplus~\cite{aflplusplus} is ``\emph{not possible without modifying the source code}''.\footnote{\url{https://aflplus.plus/docs/best_practices/\#fuzzing-a-gui-program} (as of writing).}  
Since \tool works at the abstraction of the kernel/user-mode boundary, it can fuzz GUI applications and other difficult subjects without special handling. By targeting challenging fuzzing targets, we aim to demonstrate the generality of \tool. 

In total, we collect 20 subjects as detailed in \autoref{tab:subjects}. For network protocols, we collect subjects from \profuzzbench \cite{profuzzbench}, a widely-used benchmarking platform for evaluating the net\-work-enabled fuzzers. However, for GUI applications under Linux, there is no existing fuzzing dataset. We therefore select subjects from frequently-used and well-known applications and frameworks, including text editors (UI), visual shells (UI), GNOME desktop environment (GUI), Qt (GUI), and the underlying windowing system (GUI). 

\begin{table}[t]
    \caption{Subject programs used in the evaluation.}
    \label{tab:subjects}
    \centering
    \small
    \setlength\tabcolsep{3.5pt}
    \begin{tabular}{c|lr||c|lr}
        % \toprule
        \hline
        & {\bfseries Subject} & {\bfseries Version} & & {\bfseries Subject} & {\bfseries Version} \\
        \hline
        \hline
        \multirow{10}*{\rotatebox[origin=c]{90}{\bf Network Protocols}} & DCMTK & 8326435 & \multirow{10}*{\rotatebox[origin=c]{90}{\bf GUI \& UI Applications}} & Gnome editor (gedit) & v41.0 \\
        & DNSmasq & b676923 & & Gnome Calculator & v42.9 \\
        & Exim & 5a8fc07 & & Gnome System Monitor & v42.0 \\
        & Kamailio & 2e2217b & & Glxgears & v23.0.4 \\
        & Live555 & 2c92a57 & & Midnight Commander (MC) & v4.8.27 \\
        & OpenSSH & 7cfea58 & & nano & v6.2 \\
        & OpenSSL & a7e9928 & & Vim & v8.2 \\
        & ProFTPD & 7892434 & & Wireshark & v3.6.2 \\
        & Pure-FTPd & 3296864 & & Xcalc & v1.8.6 \\
        & TinyDTLS & 0e865aa & & Xpdf & v3.04 \\
        % \bottomrule
        \hline
    \end{tabular}
    % \vspace{-10pt}
\end{table}

\subsubsection*{\textbf{Comparisons}} \label{sec:baselines}

To the best of our knowledge, no existing fuzzers target the full program environment. In the realm of fuzzing network protocols, \aflnet is the first network fuzzer, and also recommended by \aflplus for fuzzing network services. \nyxnet enhances the fuzzing throughput of \aflnet by introducing innovative hypervisor-based snapshots. Unfortunately, we cannot compare with \aflplus since it does not work on many of the network protocols. As shown in the paper of \nyxnet \cite{nyx-net}, \aflplus only works on 5 of the \profuzzbench subjects. Furthermore, for these 5 subjects, \aflplus performs significantly worse than both \aflnet and \nyxnet. Therefore, for network protocol subjects, we use \aflnet and \nyxnet as baselines for comparison.
\aflplus and \aflplus-based fuzzers are also not able to fuzz GUI applications in Linux with user interactions \cite{aflplusplus}. 
Recent work~\cite{winnie} uses {\em test harness} generation to enable GUI fuzzing, but only for Windows applications.
As such, there is no available fuzzer to compare against GUI applications under Linux. 

\subsubsection*{\textbf{Performance Metrics}} \label{sec:measures}

We evaluate the performance of \tool based on three primary metrics: \emph{bug-finding capability},  \emph{code coverage}, and \emph{fuzzing throughput}. As recommended by the fuzzing community \cite{evaluating, reliability}, the ultimate metric of a fuzzer is the number of distinct bugs found. Since a fuzzer cannot find bugs in uncovered code, code coverage is important too, and thus serves as a secondary metric. While fuzzing throughput is not a mandatory evaluation metric, it may affect the efficacy of a fuzzer.
We also report throughput to demonstrate the robustness of the fuzzer.

\begin{table*}[t]
    \setlength{\abovecaptionskip}{5pt}%    
    \caption{Statistics of bugs discovered by \toolBold; a total of 33 previously unknown bugs found, 24 bugs confirmed by developers, 16 bugs assigned CVE IDs, and 16 bugs fixed. (Note that, each color represents a distinct category of applications) }
    \label{tab:bugs}
    \centering
    % \footnotesize
    \small
    \def\arraystretch{0.96}
    \begin{tabular}{l|l|ll|ll}
        \toprule
        {\bfseries ID} & {\bfseries Subject} & {\bfseries Bug Description} & {\bfseries Environment} & {\bfseries Bug Type} & {\bfseries Bug Status} \\
        \hline
        \hline
        \rowcolor{blue!10}
        1 & Dcmtk & Failed to check bounds of stored \verb+dicom.dic+ data  & Cached data & Buffer overflow & CVE-requested, fixed\\
        \rowcolor{blue!10}
        2 & Exim &  Failed to check bounds of a corrupted \verb+resolv.conf+ & Configuration & Buffer overflow & Reported \\
        \rowcolor{blue!10}
        3  & Exim  & Glibc failed to handle an empty \verb+passwd+ line & Special file & Null pointer dereference & Reported\\
        \rowcolor{blue!10}
        4  & Kamailio & Improperly handle a corrupted client request & Socket & Null pointer dereference & Reported \\
        \rowcolor{blue!10}
        5  & Live555 & Improperly handle a malicious \verb+SETUP+ client request & Socket & Heap use after free & CVE-granted, fixed  \\
        \rowcolor{blue!10}
        6  & Live555 & Failed to check bounds of a corrupted \verb+test.mkv+ & Media resource & Buffer overflow & Reported \\
        \rowcolor{blue!10}
        7  & OpenSSH & Improperly handle a corrupted \verb+sshd_config+ & Configuration & Null pointer dereference & CVE-requested, fixed \\
        \rowcolor{blue!10}
        8  & OpenSSH & Improperly handle a corrupted \verb+gai.conf+ & Configuration & Null pointer dereference & Reported \\
        \rowcolor{blue!10}
        9  & Pure-FTPd  & Glibc failed to handle a corrupted \verb+timezone+ file & Time resource & Null pointer dereference & Reported \\

        \hline

        \rowcolor{cyan!10}
        10  & gedit  & Improperly handle a null value in \verb+parse_settings()+ & Configuration & Null pointer dereference & CVE-granted\\
        \rowcolor{cyan!10}
        11  & gedit  & Improperly handle a null value from \verb+XRRGetCrtcInfo()+ & Socket & Null pointer dereference & CVE-granted\\
        \rowcolor{cyan!10}
        12 & Calculator  & Failed to check bounds of requests, events and error IDs & Socket & Buffer overflow & CVE-granted, fixed \\
        \rowcolor{cyan!10}
        13 & Calculator  & Failed to check null value from \verb+XIQueryDevice()+ & Socket & Null pointer dereference & CVE-granted\\
        \rowcolor{cyan!10}
        14 & Calculator & Improperly handle a corrupt \verb+DBUS+ message & Socket & Null pointer dereference & CVE-requested, fixed\\
        \rowcolor{cyan!10}
        15 & Monitor & Improperly handle corrupted \verb+loaders.cache+ & Cached data & Bad free & CVE-granted\\
        \rowcolor{cyan!10}
        16 & Monitor  & Failed to handle a corrupted \verb+gtk.css+ & Theme resource & Null pointer dereference & CVE-requested, fixed\\
        \rowcolor{cyan!10}
        17 & Glxgears & Failed to check bounds of \verb+numAttribs+ in messages & Socket & Buffer overflow & CVE-granted\\
        \rowcolor{cyan!10}
        18 & Glxgears & Failed to check bounds of the string length & Socket & Buffer overflow & CVE-granted\\
        \rowcolor{cyan!10}
        19 & MC & Failed to handle a corrupted \verb+terminfo+  & Configuration & Null pointer dereference & CVE-granted\\
        \rowcolor{cyan!10}
        20 & MC  & Improperly handle a corrupted \verb+xterm-256color+ & Configuration & Arithmetic exception & CVE-granted\\
        \rowcolor{cyan!10}
        21 & MC  & Improperly process error handler of \verb+x_error_handler()+ & Socket & Null pointer dereference & CVE-granted\\
        \rowcolor{cyan!10}
        22  & nano  & Failed to handle a corrupted \verb+xterm+ file & Configuration & Null pointer dereference & Reported\\
        \rowcolor{cyan!10}
        23  & nano  & Failed to check the inconsistent directory in disk & Cached data & Null pointer dereference & CVE-granted, fixed\\
        \rowcolor{cyan!10}
        24  & Vim  & Failed to handle a corrupted \verb+xterm-256color+  & Configuration & Null pointer dereference & CVE-granted\\
        \rowcolor{cyan!10}
        25  & Vim  & Failed to handle a corrupted \verb+viminfo+ file & Cached data & Null pointer dereference & CVE-granted, fixed\\
        \rowcolor{cyan!10}
        26 & Wireshark & Failed to check null pointer in \verb+initializeAllAtoms()+ & Socket & Null pointer dereference & CVE-granted, fixed\\
        \rowcolor{cyan!10}
        27  & Xcalc  & Failed to handle null pointer from \verb+XOpenDisplay()+ & Socket & Null pointer dereference & CVE-requested, fixed\\
        \rowcolor{cyan!10}
        28 & Xcalc  & Failed to check write boundary in \verb+_XkbReadKeySyms()+ & Socket & Out-of-bounds write & CVE-granted, fixed\\
        \rowcolor{cyan!10}
        29 & Xcalc  & Failed to check read boundary in \verb+_XUpdateAtomCache()+ & Cached data & Out-of-bounds read & CVE-requested, fixed\\
        \rowcolor{cyan!10}
        30  & Xpdf  & Improperly handle invalid and corrupted locale data & Configuration & Null pointer dereference & Reported \\
        \rowcolor{cyan!10}
        31  & Xpdf  & Improperly handle invalid paper size in configuration & Configuration & Null pointer dereference & Reported\\
        \rowcolor{cyan!10}
        32  & Xpdf & Failed to check pointer boundary returned from response & Socket & Bad free & CVE-requested, fixed\\
        \rowcolor{cyan!10}
        33  & Xpdf & Failed to check array boundary returned from X server & Socket & Out-of-bounds read & CVE-requested, fixed\\

        \bottomrule
    \end{tabular}
    \vspace{-5pt}
\end{table*}

\subsubsection*{\textbf{Experimental Infrastructure}} \label{sec:infrastructure}
All experiments were conducted on an Intel{\textregistered} Xeon{\textregistered} Platinum 8468V CPU with 192 logical cores clocked at 2.70GHz, 512GB of memory, and running Ubuntu 22.04.3 LTS. Each experiment runs for 24 hours.  We report the average over 10 runs to mitigate the impact of randomness.

\subsection{Discovering New Bugs (RQ.1)} \label{sec:new_Bugs}

\subsubsection*{\textbf{Method}} We ran \tool on the subjects listed in \autoref{tab:subjects} to discover bugs. We utilized the same bug oracles as traditional fuzzers (e.g., \aflnet and \nyxnet), including crashes, hangs, assertion failures, and sanitizer violations. For initiating the fuzz campaign, we used initial seeds provided by the programs if available; otherwise, we provided standard user inputs as initial seeds. In the case of network protocols, we utilized their clients to send request messages. For GUI applications, we simulated typical user interactions; as an example, with a calculator, the application is opened to perform a simple addition calculation before it is closed. All inputs represent normal usage scenarios encountered in the real world. We subjected each program to a {\em 24-hour run} (typical recommended length of a fuzz campaign \cite{evaluating}) to identify bugs. Upon finding bugs, we reported them to the developers for confirmation. In the case of bugs with potential security implications, we requested CVE IDs from the CVE Numbering Authority. All activities were conducted in a one-month period, including bug finding, debugging, reporting to developers, and requesting CVEs.

\subsubsection*{\textbf{Results}} \autoref{tab:bugs} shows the distinct and previously unknown bugs found by \tool. In the \emph{Bug Description} column, we elucidate the root causes responsible for these bugs, and illustrate the immediate environmental factors in the \emph{Environment} column. It is important to note that triggering a bug often requires hundreds of diverse environmental inputs. Therefore, we only listed the most relevant environmental input that exposed the bugs after mutation. Furthermore, we provide details about the bug types and their current status in the last two columns.

In total, we discovered 33 previously unknown bugs, out of which 24 have received confirmation from their respective developers. Developers had fixed 16 of these bugs by the time of paper submission.  16/24 bugs have been assigned CVE IDs. These bugs span various categories, including buffer overflows, use-after-frees, null pointer dereferences, and arithmetic exceptions. Furthermore, these bugs are triggered by fuzzing a diverse range of environmental inputs, including sockets, configuration files, multiple types of resource files, cached data, etc. Therefore, a fuzzer that exclusively concentrates on a singular input cannot expose all of these bugs.   

These results highlight the significant bug-finding capability of \tool. Moreover, they demonstrate the importance of program environment fuzzing, and \tool has shown its effectiveness in this regard. 
Two case studies below illustrate bugs discovered by \tool. 

\subsubsection*{\textbf{Case study: GNOME Desktop Environment}}
GNOME client applications (e.g., \verb+gnome-calculator+, etc.) interact with the windowing system and several other services (\autoref{fig:calc}).
\tool is able to expose several bugs in multiple different input sources, including several bugs related to the windowing system and client libraries, bugs in the DBus socket connection to the session manager, as well as bugs in non-socket inputs (\verb+loaders.cache+, \verb+gtk.css+, etc.).
As an example, we can consider Bug \#12, which affects the \verb+XESetWireToEvent()+ function from \verb+libX11+.
This function fails to check whether the event values are within the bounds of the arrays that the functions write to. Instead, the function directly uses the value as an array index, leading to an intra-object overwrite and probable crash. This bug stems from the implicit trust that \verb+libX11+ functions place in the values supplied by an X server, following X11 protocol. However, the environment cannot be fully trusted, as a malicious server or proxy can impact applications. This bug was assigned CVE ID by X11 developers and received a CVSS score HIGH 7.5. We note that other subjects, including many GNOME applications, are also affected by this bug.  

\subsubsection*{\textbf{Case study: Bug \#23 in GNU nano}} GNU nano is a text editor for Unix-like operating systems and is part of the GNU Project. This bug appears in \verb+read_the_list()+ of \verb+browser.c+. This function initiates an initial iteration over a directory using \verb+readdir()+ to obtain the current entries, followed by a rewinding action using \verb+rewinddir()+ to cache these entries. Subsequently, a second iteration employing \verb+readdir()+ is performed to directly access these cached entries. Unfortunately, before this second iteration, there is no boundary-checking mechanism. As a result, any environmental changes, such as directory deletions, can easily trigger a crash during the second iteration. This is precisely how \tool exposes it. This bug existed from the first version of GNU nano in 2005 and had been hidden for 18 years! 

\result{\tool discovered 33 previously unknown bugs in widely used network protocols and GUI applications, with 24 confirmed and 16 fixed by their developers. 16 of them were assigned CVEs.}

\subsection{Comparisons with Baselines (RQ.2)} \label{sec:comparison}

\subsubsection*{\textbf{Method}} For network protocols, we compare \tool against two baselines \aflnet and \nyxnet under three aspects: the number of bugs found, code coverage, and fuzzing throughput. We omit GUI programs due to the lack of a suitable baseline. We configure all fuzzers employing the same initial seeds obtained from \profuzzbench. Our evaluation of code coverage focuses on measuring branch coverage achieved on binaries. We utilize the original scripts provided by \profuzzbench, to collect code coverage data and present their trends over time. We report the total number of bugs found, the average coverage, and the average fuzzing throughput achieved by each fuzzer across 10 runs of 24 hours.

\subsubsection*{\textbf{Comparing Results on Bug Finding.}} \autoref{tab:comparing_bugs} shows the total number of unique bugs found by each fuzzer. In all subjects, \tool discovered a total of 9 unique bugs, as detailed in \autoref{tab:bugs}. However, both \aflnet and \nyxnet could only find 2 (\ie Bug \#4 and Bug \#5 in \autoref{tab:bugs}); in addition, neither fuzzer found any additional bug. The remaining 7 bugs were exposed by fuzzing non-socket environment inputs, such as cached data and resources. Since these environment inputs are not fuzzing targets for \aflnet and \nyxnet, they were unable to expose them. Furthermore, regarding bugs induced by network sockets, \tool successfully exposed the same number as \aflnet and \nyxnet. This demonstrates that \tool maintains the effectiveness in fuzzing a single environment source, although it fuzzes all environment sources.

\result{In the aspect of bug finding, \tool discovered 9 previously unknown bugs, while \aflnet and \nyxnet only discovered 2 without any additional bug found. }

\begin{table}[t]
    \setlength{\abovecaptionskip}{5pt}% 
    \centering
    \small
    \caption{Number of unique bugs found by \aflnetBold, \nyxnetBold and \toolBold on subjects of network protocols.}
    \label{tab:comparing_bugs}
    \setlength\tabcolsep{8pt}
    \begin{tabular}{l|rrr}
        \toprule
        {\bfseries Fuzzer} & {\aflnetBold} & {\nyxnetBold} & {\toolBold} \\
        \hline
        \hline
        {\bfseries \#Bug} & 2 & 2 & 9 \\
        \bottomrule
    \end{tabular}
    \vspace{-.3cm}
\end{table}

\subsubsection*{\textbf{Comparing Results on Code Coverage}} 
\begin{figure*}[h]
\setlength{\abovecaptionskip}{5pt}
  \centering
  \includegraphics[page=1, trim= 1.15in 2.92in 1.15in 0.57in, clip, scale=0.64]{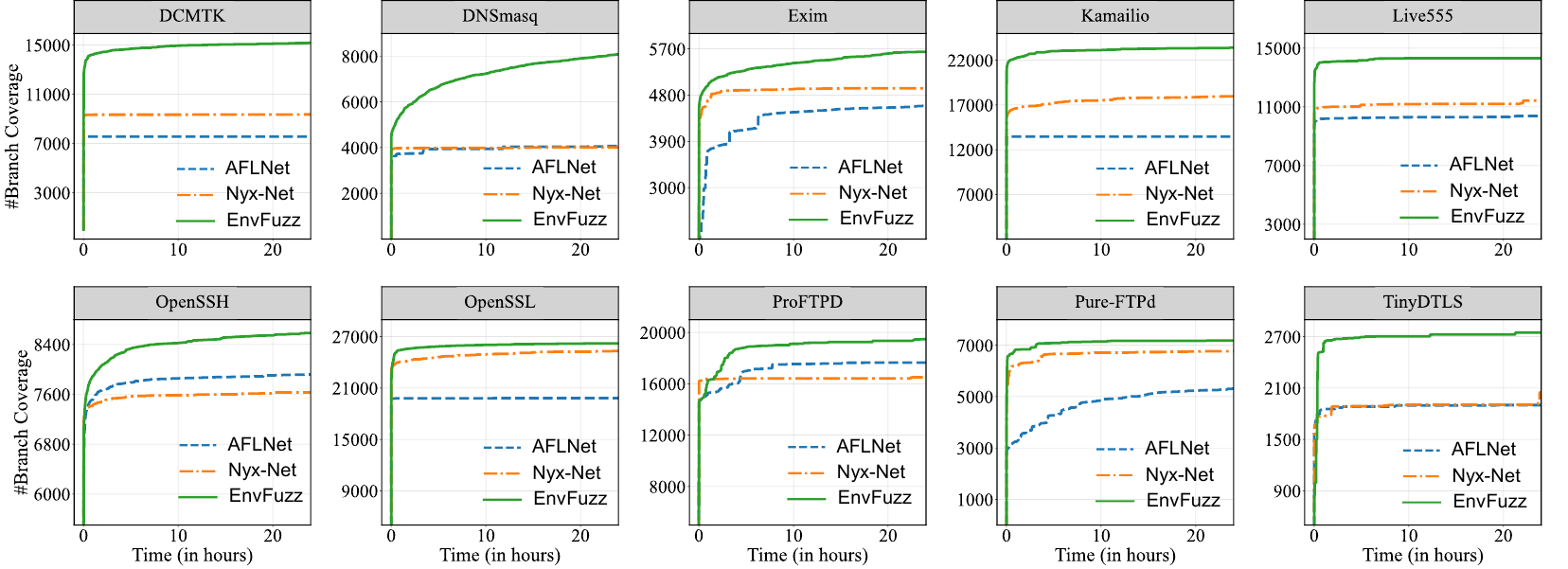}
  \caption{ Code covered over time by \aflnetBold, \nyxnetBold and \toolBold across 10 runs of 24 hours on \profuzzbenchBold subjects.}
  \label{fig:coverage-trend}
\end{figure*}

\autoref{fig:coverage-trend} illustrates trends in average code coverage over time for \aflnet, \nyxnet and \tool. Across all subjects, \tool significantly outperformed both \aflnet and \nyxnet. Initially, at the start of each experiment, all three fuzzers covered a similar number of code branches. However, over time, \tool substantially covered more code than \aflnet and \nyxnet. Even after 24 hours, \tool still had the potential to discover new code, whereas, in most cases, the code coverage for \aflnet and \nyxnet tended to plateau quickly.

\begin{table}[t]
    \setlength{\abovecaptionskip}{5pt}%    
    \caption{Average branch coverage across 10 runs of 24 hours achieved by \toolBold compared to \aflnetBold and \nyxnetBold. }
    \label{tab:coverage_results}
    \centering
    \small
    \setlength\tabcolsep{2.2pt}
    \begin{tabular}{l|r|rrr|rrr}
        \toprule
        \multirow{2}{*}{\bfseries Subject} & \multirow{2}{*}{\bfseries \toolBold} & \multicolumn{3}{c|}{ \bfseries \footnotesize Compare with \aflnetBold} &  \multicolumn{3}{c}{ \bfseries \footnotesize Compare with \nyxnetBold}  \\ 
        \cline{3-8} 
         & & {\bfseries \footnotesize Coverage} & { \bfseries \footnotesize Improv} &  {\footnotesize ${\boldsymbol{\hat{A}_{12}}}$ } &  {\bfseries \footnotesize Coverage} & { \bfseries \footnotesize Improv} &   {\footnotesize ${\boldsymbol{\hat{A}_{12}}}$ } \\
        \hline
        \hline
        DCMTK & 15181.7 & 7564.9 & +100.69\% & 1.00 & 9362.0 & +62.16\% & 1.00 \\
        DNSmasq & 8090.9 & 4066.7 &  +98.95\% & 1.00 & 4009.0 & +101.82\% & 1.00 \\
        Exim & 5642.7 & 4594.4 &	+22.82\%	& 1.00 &	4935.2	& +14.34\%	& 1.00	\\
        Kamailio & 23425.6 & 13466.1 & +73.96\% & 1.00 & 17960.0 & +30.43\% & 1.00 \\
        Live555 & 14319.0 & 10379.5 & +37.95\% & 1.00 & 11436.0 & +25.21\% & 1.00 \\
        OpenSSH & 8584.5 & 7920.0	& +8.39\%	& 1.00 &	7631.5	& +12.49\%	&	1.00 \\
        OpenSSL & 26225.9 & 19820.4 & +32.32\% & 1.00 & 25330.1 & +3.54\% & 1.00 \\
        ProFTPD & 19478.0 & 17654.0 & +10.33\% & 1.00 & 16504.0 & +18.02\% & 1.00 \\
        Pure-FTPd & 7182.8 & 5309.0 & +35.29\% & 1.00 & 6766.5 & +6.15\% & 1.00 \\
        TinyDTLS & 2747.5 & 1901.5 & +44.49\% & 1.00 & 2052.5 & +33.86\% & 1.00 \\
        \hline
        \hline
        \bfseries Average & - & - & +46.52\% & - & - & +30.80\% & - \\
        \bottomrule
    \end{tabular}
    % \vspace{-10pt}
\end{table}

\autoref{tab:coverage_results} shows the  final code coverage of \tool and two baselines. To quantify the improvement of \tool over baselines, we report the number of branches covered by \tool, \aflnet and \nyxnet (\emph{Coverage}), respectively, the percentage improvement of \tool (\emph{Improv}), and the probability that a random campaign of \tool outperforms a random campaign of baselines (${\hat{A}_{12}}$). 
For all subjects, \tool covers more code than both baselines. Specifically, \tool averagely covers 46.52\% more code than \aflnet with a range from 8.39\% to 100.69\%. When compared to \nyxnet, \tool covers 30.80\% more code on average from 3.54\% to 101.82\%. The Vargha-Delaney \cite{vargha-delaney} effect size ${\hat{A}_{12}} \geq 0.70$ demonstrates a substantial improvement of \tool over both baselines in terms of code coverage.

To investigate the correlation between improved code coverage and program environment fuzzing, we conducted a comprehensive analysis of the additional code covered by \tool, focusing on the subject DCMTK. DCMTK is a widely-used implementation of the DICOM (Digital Imaging and Communication in Medicine) protocol.
While fuzzing DCMTK using \tool, we observed multiple environment sources that undergo mutation. These included the configuration file, the database responsible for storing patient records, various patient cases, and network sockets utilized for hospital communication. Among the 5819 additionally covered branches, 69\% of them demonstrated direct connections to environmental mutations, such as parsing and changing the configuration settings and adding entries to the database. Therefore, full environment fuzzing significantly contributes to increased code coverage.

\result{\tool covers 46.52\% and 30.80\% more code than \aflnet and \nyxnet, respectively, with most additional code coverage resulting from program environment fuzzing. }

\begin{table}[t]
    \setlength{\abovecaptionskip}{3pt}%    
    \setlength{\belowcaptionskip}{0pt}%
    \caption{Fuzzing throughput (execs/s) in 10 runs of 24 hours achieved by \toolBold compared to \aflnetBold and \nyxnetBold. }
    \label{tab:throughput}
    \centering
    \small
    \setlength\tabcolsep{3.5pt}
    \begin{tabular}{l|r|rr|rr}
        \toprule
        \multirow{2}{*}{\footnotesize \bfseries Subject} & \multirow{2}{*}{\footnotesize \bfseries \toolBold} & \multicolumn{2}{c|}{\footnotesize \bfseries Compare with \aflnetBold} &  \multicolumn{2}{c}{ \footnotesize \bfseries Compare with \nyxnetBold}  \\ 
        \cline{3-6} 
         & & {\footnotesize \bfseries \aflnetBold} & {\footnotesize \bfseries Speedup} &  {\footnotesize \bfseries \nyxnetBold} & {\footnotesize \bfseries  Speedup}  \\
        \hline
        \hline
        DCMTK     & 101.7  & 22.3 & 4.57$\times$ & 815.4 & 0.12$\times$ \\
        DNSmasq   & 393.0 & 22.6 & 17.40$\times$& 1126.8 & 0.35$\times$ \\
        Exim      & 713.4	& 5.1 &	139.06$\times$ & 514.5 & 1.39$\times$ \\
        Kamailio  & 121.9  & 5.2  & 23.62$\times$& 234.9  & 0.52$\times$ \\
        Live555   & 237.4 & 16.8 & 14.13$\times$ & 133.1 & 1.78$\times$\\
        OpenSSH   & 1320.9	& 38.6	& 34.19$\times$ & 1031.1	& 1.28$\times$ \\
        OpenSSL   & 124.5 & 32.7 & 3.80$\times$ & 227.4  & 0.55$\times$ \\
        ProFTPD   & 293.4 & 7.2  & 40.91$\times$& 333.5  & 0.88$\times$ \\
        Pure-FTPd & 528.9 & 9.8  & 54.08$\times$& 596.0 & 0.89$\times$ \\
        TinyDTLS  & 640.9 & 3.1  & 206.74$\times$& 1354.0 & 0.47$\times$\\
        \hline
        \hline
        \bfseries Average & -	& - & 53.85$\times$ & 	-	& 0.82$\times$ \\
        \bottomrule
    \end{tabular}
    \vspace{-10pt}
\end{table}

\subsubsection*{\textbf{Comparing Results on Fuzzing Throughput}} The experimental results on fuzzing throughput are shown in \autoref{tab:throughput}. For each fuzzer, the corresponding fuzzing throughput is shown in the respective columns. In the \emph{Speedup} columns, we present how much faster \tool executes compared to \aflnet and \nyxnet, respectively.  \tool achieves a fuzzing throughput ranging from 101.7 to 1320.9 executions per second. The fuzzing throughput of \aflnet is from 3.1 to 38.6 executions per second, and \tool executes 53.85 $\times$ faster than \aflnet on average. When compared to \nyxnet, \tool executes faster than \nyxnet on some subjects (e.g., 1.78$\times$ faster on Live555) but slower on others. These results are expected as \aflnet always replays each input sequence from the root, while both \tool and \nyxnet avoid replaying repetitive input sequences by faithful replay and state snapshots, respectively. In addition, compared to \nyxnet, \tool introduces some time overhead on certain subjects (e.g., DCMTK) to explore more behaviors. However, this overhead is justified by the evident improvement in bug finding and code coverage. Overall, \tool still maintains a robust throughput.

\result{\tool maintains a robust fuzzing throughput while enhancing the capability of bug finding and code coverage.}

\subsection{Ablation Studies (RQ.3)} \label{sec:ablation}

\subsubsection*{\textbf{Impact of Algorithm Components}} \tool employs two strategies to enhance the search efficiency of the program environment: behavior divergence handling based on the relaxed replay, and feedback guidance. To evaluate the impact of each strategy on the improvement of the code coverage, we conducted an ablation study. For this purpose, we developed two ablation tools:
\begin{itemize}[leftmargin=*]
\item \efone: based on \tool, without behavior divergence handling,
\item \eftwo: based on \tool, without fuzzing feedback.
\end{itemize}
We compare the average code coverage achieved by \tool with that of \efone and \eftwo across 10 runs of 24 hours in each subject, and report the percentage improvements. 

\begin{table}[t]
    \setlength{\abovecaptionskip}{5pt}%    
    \caption{Improvement of code coverage achieved by \toolBold in comparison to ablation tools \efoneBold and \eftwoBold.
    The results show that the impact of behavior divergence handling and fuzzing feedback is significant.}
    \label{tab:ablation}
    \centering
    \small
    \begin{tabular}{l|rr||l|rr}
        \toprule
         {\bfseries Subject} & {\bfseries \textit{vs.} \efoneBold} & {\bfseries \textit{vs.} \eftwoBold} & {\bfseries Subject} & {\bfseries \textit{vs.} \efoneBold} & {\bfseries \textit{vs.} \eftwoBold} \\
        \hline
        \hline
        DCMTK & +60.83\% & +22.52\% & gedit & +22.14\% & +8.17\% \\
        DNSmasq & +39.28\% & +27.79\%  & Calculator & +27.12\% & +6.61\% \\
        Exim &	+12.24\% & +9.66\%	 & Monitor & +14.24\%	& +4.44\%	\\
        Kamailio & +28.89\% & +10.52\%  & Glxgears & +12.01\% & +2.39\%\\
        Live555 & +30.26\% & +14.61\%  & MC & +68.60\% & +13.46\% \\
        OpenSSH & +10.92\% & +3.99\%	& nano & +20.48\%	&	+8.75\% \\
        OpenSSL & +12.98\% & +8.06\%  & Vim &  +12.50\% & +20.47\% \\
        ProFTPD & +26.81\% & +9.21\%  & Wireshark & +17.90\% & +8.17\% \\
        Pure-FTPd & +46.21\% & +6.75\%  & Xcalc & +27.66\% & +5.55\% \\
        TinyDTLS& +98.57\% & +8.59\%  & Xpdf & +22.19\% & +7.79\%\\
        \hline
        \hline
        \multicolumn{4}{c|}{\bfseries Average} & +30.59\%  & +10.38\% \\
        \bottomrule
    \end{tabular}
    \vspace{-10pt}
\end{table} 

\autoref{tab:ablation} shows the results of the percentage improvements in terms of average code coverage. Overall, across all subjects, both strategies contributed to the increase in code coverage, with none exhibiting a negative impact. Compared to \efone without behavior divergence handling, \tool resulted in an average increase of 30.59\% in code coverage. Notably, in DCMTK, TinyDTLS and MC, \tool exhibited code coverage improvements exceeding 60\%. Compared to \eftwo without fuzzing feedback, \tool increased the code coverage by 2.39\% to 27.79\%, with an average increase of 10.38\%. Furthermore, comparing \tool with both tools across all subjects, ${\hat{A}_{12}}{=}1$, which indicates that \tool significantly outperforms \efone and \eftwo. These results demonstrate the importance of \tool's divergence handling and the effectiveness of fuzzing feedback in guiding the search. 

We further measured the fuzzing throughput of \tool, \efone, and \eftwo across each subject. On average, \tool achieves a fuzzing throughput of 447.6 executions per second, while \eftwo achieves a similar throughput of 454.9 executions per second. However, \efone executes faster than \tool with a fuzzing throughput of 698.3 executions per second. This higher throughput is due to \tool's strategy of handling behavior divergence, which explores longer traces, trading raw throughput for better coverage.

\result{ Divergence handling and feedback guidance enable \tool to increase code coverage by 30.59\% and 10.38\%, respectively. The contribution of each strategy to enhancing code coverage is significant. }

\subsubsection*{\textbf{Analysis of Relaxed Replay}} To further analyze the impact of relaxed replay for divergence handling, we examined the following additional questions:
\begin{itemize}[leftmargin=*]
\item How often do executions resort to relaxed replay (\#Freq.)?
\item How many system calls in tree branches use relaxed replay (\#RelaxSysCs \emph{vs} \#TotalSysCs)?
\item How early after a branch does relaxed replay start (\#StartPoint)?
\end{itemize}
For this purpose, we collect the statistical data from 20 subjects over 24-hour runs and report them in \autoref{tab:relaxed_replay}. On average, 84.48\% of all executions across the 20 subjects have to resort to relaxed replay. The total number of system calls executed in each tree branch (calculated from forking points) is 187.9, with 108.1 of those system calls using relaxed replay. In addition, the starting points of the relaxed replay vary among different subjects, ranging from 5.46\% of the tree branch on TinyDTLS to 89.87\% on OpenSSH. On average, the relaxed replay starts to resort to relaxed replay around halfway (49.97\%). These results demonstrate that the relaxed replay for divergence handling is necessary and commonly used by \tool.

\begin{table}[t]
    \setlength{\abovecaptionskip}{5pt}%    
    \caption{Statistical analysis of relaxed replay proposed by \toolBold, including the frequency of the executions resorting to relaxed replay (\#Freq.), the total number of system calls executed in each tree branch (\#TotalSysCs), the number of system calls resorting to relaxed replay in each tree branch (\#RelaxSysCs), and the point at which a tree branch starts to resort to relaxed replay (\#StartPoint).}
    \label{tab:relaxed_replay}
    \centering
    \small
    \setlength\tabcolsep{4pt}
    \begin{tabular}{l|r|rrr}
        \toprule
         {\bfseries Subject} & {\bfseries \#Freq.} & {\bfseries \#TotalSysCs} & {\bfseries \#RelaxSysCs} & {\bfseries \#StartPoint}\\
        \hline
        \hline
        DCMTK &  93.67\%	& 130.0 &	116.7	&10.27\% \\
        DNSmasq & 34.52\%	& 99.1	& 11.1	& 88.82\%   \\
        Exim &	 98.78\%	& 90.8	& 23.4	& 74.18\%	\\
        Kamailio & 73.85\%	& 141.2	& 90.5	& 35.93\% \\
        Live555 &  45.12\%	& 375.1	& 130.2	& 65.30\% \\
        OpenSSH &  87.66\%	& 105.3	& 10.7	& 89.87\%\\
        OpenSSL &  91.02\%	& 51.3	& 11.0	& 78.57\%\\
        ProFTPD &  95.01\%	& 172.1	& 24.8	& 85.61\% \\
        Pure-FTPd & 60.08\%	& 114.9	& 14.5	& 87.41\%  \\
        TinyDTLS& 78.58\%	& 312.6	& 295.5	& 5.46\%\\
        \hline
        gedit &  99.89\% & 397.3 & 332.1 & 16.42\% \\
        Calculator & 99.65\% & 152.1 & 80.7 & 46.98\% \\
        Monitor & 94.68\% & 114.5 & 69.6 & 39.21\% \\
        Glxgears & 94.31\%  & 92.6 & 35.8 & 61.37\%  \\
        MC & 89.98\% & 278.1 & 127.6 & 54.11\%  \\
        nano & 98.99\% & 164.6 & 89.0 & 45.93\%  \\
        Vim & 93.69\% & 386.2 & 307.7 & 20.32\%  \\
        Wireshark & 81.43\% & 198.4 & 157.5 & 20.60\% \\
        Xcalc & 89.37\% & 147.5 & 51.7 & 50.78\% \\
         Xpdf & 89.30\%  & 234.1 & 182.0 & 22.26\% \\
        \hline
        \hline
        {\bfseries Average} & 84.48\%	& 187.9	& 108.1	& 49.97\% \\
        \bottomrule
    \end{tabular}
    \vspace{-10pt}
\end{table} 

\subsection{Discussion}

\subsubsection*{\textbf{Manual Effort}}

The manual effort needed for using \tool is minimal. The only manual involvement is the user inputs necessary for testing GUI applications in the recording phase. For example, when testing the calculator, the user needs to open the application, execute a simple addition operation, and then close it. After recording, the rest of the fuzzing workflow is fully automatic.
To collect code coverage feedback, \tool can directly instrument the binaries of the program under test, eliminating the need to recompile from source code.
Similarly, the system-call interception infrastructure (for record and replay) is designed to work with binary code.

\subsubsection*{\textbf{Limitations}}

In this paper, we leverage greybox fuzzing over complex program environments. We have demonstrated that the approach of \tool is effective in exposing previously unknown bugs and enhancing code coverage. Like other fuzzers, the efficacy of \tool depends on the quality of the initial seed(s), and \tool is not guaranteed to cover the entire search space.
Limited seed recordings (\eg open and immediately close a GUI) generally result in a limited exploration compared to diverse recordings (\eg exercising different GUI elements).
However, the dependence on quality seeds is an inherent limitation of fuzzing in general, in contrast to model checking and other verification techniques that attempt to systematically explore the entire search space.
While \tool can fuzz a broad range of programs, its scope is limited to Linux user-mode environments. This limitation stems from our underlying environmental record and replay infrastructure.
Despite this, and compared to existing fuzzers, \tool still maintains its generality, and can fuzz even challenging subjects such as network protocols and GUI applications. 

Relaxed replay assumes that I/O system calls can be mutated and reordered arbitrarily.
This is a straightforward generalization of what existing fuzzers already assume. For example, \afl implicitly assumes the input file can be mutated arbitrarily, while \aflnet assumes messages can be reordered. However, these assumptions may not always hold for some edge cases. Special files (\eg \verb+/proc/*+ and \verb+/dev/zero+) and self-pipes are not mutable.
Fortunately, such examples are rare and can be avoided using a predefined special-case list. 
As such, no false positives were detected during our evaluation.
By design, \tool does not use modeling, allowing it to fuzz programs without any manual effort or prior knowledge. In addition, \tool supports fuzzing ``new'' syscalls not present in the original recording. During relaxed replay, \tool provides inputs to the syscall based on file descriptors, regardless of the specific syscall number. However, due to behavior divergence, if the program invokes a system call to access inputs from a file descriptor that is not originally recorded, \tool will fail the system call to maintain the plausibility of the replay.

\section{Related Work} \label{sec:related_work}

\subsubsection*{\textbf{Environment Capture}} 
Environment handling poses a critical challenge in the realm of model checking and symbolic execution, where achieving an accurate analysis of program behaviors requires considering the full surrounding environment. Many existing approaches manually abstract the environment via a model \cite{ball2006thorough, klee, dart, cute, cmc}, but crafting abstract models is labor-intensive. Some alternatives \cite{s2e, poeplau2020symbolic} leverage virtualization to eliminate the need for constructing abstract models. However, the path-explosion problem persists when analyzing an entire software stack \cite{baldoni2018survey, s2e}; the presence of many program environments further exacerbates the path-explosion problem while finding bugs in software.  

\subsubsection*{\textbf{Fuzzing Effort}}
In the area of fuzzing, existing fuzzers often focus on a single input, disregarding other environment sources. The potential solutions for capturing environment effects, are the use of Virtual Machine (VM) fuzzing \cite{nyx-net} allowing the target to be fuzzed in the context of an emulated system environment, or overriding \texttt{glibc} functions \cite{mirzamomen2023finding}. VM-based fuzzing is a heavyweight solution. Moreover, both approaches cannot hook the full environment, which misses environmental interactions with external servers, hardware devices, and human users. In contrast, our approach is lightweight yet robust, effectively handling the full environment.

\subsubsection*{\textbf{Stateful Fuzzing}} Many programs are stateful, processing inputs based on their internal states. 
While fuzzing stateful programs, relying solely on code coverage is insufficient in guiding fuzzers to explore complex state machines and reach deep states \cite{ijon, aflnet, nsfuzz, stateful, chatafl}. Identifying program states poses a significant challenge, and several works propose diverse state representation schemes. \ijon \cite{ijon} uses human code annotations to annotate states, and \aflnet \cite{aflnet} manually extracts response code based on network protocols as states (\eg 404 for \texttt{http}). \stateafl \cite{stateafl} hashes in-memory variables as states, while \sgfuzz \cite{stateful} and \nsfuzz \cite{nsfuzz} utilize enum variables as states with manual filtering. However, these approaches involve much manual effort or employ specific heuristics such as the emphasis on enum variables in \sgfuzz.

\subsubsection*{\textbf{Snapshot Fuzzing}} When fuzzing stateful systems, achieving a deep exploration of program states often requires a lengthy sequence of messages. For instance, \aflnet \cite{aflnet} opts for replaying each message sequence from initial states, somewhat impeding its fuzzing speed. To address this limitation, \snapfuzz \cite{snapfuzz} employs an in-memory filesystem to efficiently reset to specific interesting states, overcoming the impediment faced by \aflnet. In a similar vein, \nyxnet \cite{nyx-net} introduces a hypervisor-based technique to dump program states at points of interest, including all memory contents. 
Our algorithm eliminates the need for snapshots or hypervisors, and dynamically {\em reconstructs} states on-demand through replay. Our algorithm has similarities with \verb+fork+-based fuzzers such as \afl \cite{afl} and \aflplus \cite{afl++}. Rather than employing a global fork server at program entry, we implement a mini-fork server at each program input, avoiding replaying system call sequence prefixes.

\subsubsection*{\textbf{Record and Replay}} Record and replay have been widely used in assisting program analysis \cite{rrdebug, revirt, repeatable, vidi}. These approaches have targeted different software and hardware, including virtual machines \cite{revirt, repeatable}, user-space programs \cite{rrdebug, sparse_replay} and hardware \cite{vidi, gpureplay}. Among them, \texttt{rr} \cite{rrdebug} is a well-known debugger for its ease of use and low adoption cost. Its design principle is to record and replay unmodified user-space applications (binaries) with stock Linux kernels, with a fully user-space implementation running without special privileges, and without using pervasive code instrumentation. The \texttt{rr} debugger is primarily designed to help with difficult-to-reproduce bugs that depend on nondeterministic elements of the environment. Our approach has some similarities with the \texttt{rr} debugger, but we re-purpose \texttt{rr} debug-style replay for bug discovery by employing relaxed replay. 
\section{Conclusion} \label{sec:conclusion}

In this paper, we have proposed a methodology, tool, and evaluation
to handle complex program environments. Our \tool tool avoids environment modeling by recording program executions and selectively mutating (in the style of greybox fuzzing) the recorded executions during replay to capture the effect of different environments. Evaluation of \tool found 33 previously unknown bugs, out of which 24 were confirmed by developers. The applications tested include well-known GUI applications and protocol implementations.
\tool presents a general approach for handling software environments, which is different from (a) the practitioners' approach of procuring sample environments for testing code on them one by one, or (b) the current established research on environment modeling.  We do not model environments and we do not procure environments. Instead \tool is an automated framework for {\em implicitly} navigating the space of program environments via mutational fuzzing.

\section*{Acknowledgment}
We thank the anonymous reviewers for their suggestions. We also thank Van-Thuan Pham for his comments on the draft. This research is supported by the National Research Foundation, Singapore, and Cyber Security Agency of Singapore under its National Cybersecurity R\&D Programme (Fuzz Testing NRF-NCR25-Fuzz-0001). Any opinions, findings and conclusions, or recommendations expressed in this material are those of the author(s) and do not reflect the views of National Research Foundation, Singapore, and Cyber Security Agency of Singapore.

\bibliographystyle{plain}
\balance
\bibliography{references}

\end{document}